\documentclass[a4paper,11pt]{article}

\usepackage{amsmath}
\usepackage{amssymb}
\usepackage{epsfig}

\newcommand{\ord}[1]{\mathcal{O}\left( #1 \right)}
\newcommand{\Fig}[1]{Fig.~\ref{fig:#1}}
\newcommand{\fig}[1]{fig.~\ref{fig:#1}}
\newcommand{\Figs}[1]{Figs.~\ref{fig:#1}}
\newcommand{\figs}[1]{figs.~\ref{fig:#1}}
\newcommand{\Eq}[1]{Eq.~(\ref{eq:#1})}
\newcommand{\eq}[1]{eq.~(\ref{eq:#1})}

\newcommand{\eqs}[1]{eqs.~(\ref{eq:#1})}

\newcommand{\Eqi}[2]{Eq.~(\ref{eq:#1}#2)}
\newcommand{\eqi}[2]{eq.~(\ref{eq:#1}#2)}

\newcommand{\MeV}{\,\mathrm{MeV}}
\newcommand{\keV}{\,\mathrm{keV}}
\newcommand{\eV}{\,\mathrm{eV}}
\newcommand{\cm}{\,\mathrm{cm}}
\newcommand{\erg}{\,\mathrm{erg}}

\newcommand{\vev}[1]{\left\langle #1\right\rangle}
\newcommand{\dm}[1]{{\Delta m^2_{#1}}}
\newcommand{\fracwithdelims}[4]{\left#1 \frac{#3}{#4} \right#2}

\newcommand{\capdef}{}
\newcommand{\mycaption}[2][\capdef]{\renewcommand{\capdef}{#2}%
        \caption[#1]{{\itshape #2}}} 
\makeatletter
\renewcommand{\fnum@table}{\textbf{\tablename~\thetable}}
\renewcommand{\fnum@figure}{\textbf{\figurename~\thefigure}}
\makeatother

\newlength{\myem}
\newcommand{\sep}[1]{#1}
\newcounter{mysubequation}[equation]
\renewcommand{\themysubequation}{\alph{mysubequation}}
\newcommand{\mytag}{\stepcounter{mysubequation}%
\tag{\theequation\protect\sep{\themysubequation}}}
\newcommand{\globallabel}[1]{\refstepcounter{equation}\label{#1}}

\newcommand{\td}{t_{\text{diff}}}
\newcommand{\tn}{t_{\nu}}
\newcommand{\tnb}{t_{\bar\nu}}
\newcommand{\Y}{Y_{L}}
\newcommand{\dY}{\delta Y_{L}}
\newcommand{\Ya}{Y^0_{L}}
\newcommand{\mmR}{m^2\! R}

\newcommand{\nB}{n_{B}}
\newcommand{\mun}{\mu_\nu}

\newcommand{\GF}{G_{\text{F}}}
\newcommand{\SN}{SN 1987A}
\newcommand{\km}{\,\mathrm{km}}
\renewcommand{\sec}{\,\mathrm{sec}}
\newcommand{\gcm}{\,\mathrm{g}\,\mathrm{cm}^{-3}}
\newcommand{\Msun}{M_{\odot}}
\newcommand{\MP}{M_{\text{Pl}}}
\newcommand{\Mf}{M_*}
\newcommand{\meff}{m_{\text{eff}}}
\newcommand{\Ploss}{P(\nu_e\to\text{bulk})}
\newcommand{\Plossbar}{P(\bar\nu_e\to\text{bulk})}

\newcommand{\myskip}{\medskip}
\DeclareMathOperator{\sign}{sign}

\newcommand{\SNS}{Scuola Normale Superiore and INFN, Sezione di Pisa, \\
I--56126 Pisa, Italy}

\newcommand{\titletext}{Bulk neutrinos and core collapse supernovae} 
\newcommand{\authortext}{\large G. Cacciapaglia, M. Cirelli, Y. Lin,
  A. Romanino
  \medskip\\\em\normalsize \SNS} 
\newcommand{\abstracttext}{We discuss the phenomenology of neutrino
  mixing with bulk fermions in the context of supernova physics. The
  constraints on the parameter space following from the
  usual energy loss argument can be relaxed by four orders of
  magnitude due to a feedback mechanism that takes place in a broad
  region of the parameter space. Such a mechanism also affects the
  protoneutron star evolution through a non trivial interplay with
  neutrino diffusion.  The consistency with the \SN\ signal is
  discussed, as well as the implications for deleptonization, cooling,
  composition of the neutrino flux and the delayed explosion
  scenario.}

\title{
\normalsize
\LARGE\bfseries\titletext\bigskip}
\author{\begin{minipage}[t]{0.8\textwidth}
\normalsize\centering\authortext
\end{minipage}}
\date{}

\begin{document}

\bigskip
\maketitle
\begin{abstract}\normalsize\noindent
\abstracttext
\end{abstract}\normalsize\vspace{\baselineskip}


\noindent

\section{Introduction}

Sterile neutrinos from extra dimensions are not likely to play a
primary role in atmospheric and solar neutrinos~\cite{sno}. On the
contrary, astrophysics, cosmology and rare decays still represent
natural stages for their exotic performances. In this paper, we focus
on the implications of a possible mixing of the electron neutrino with
a Kaluza-Klein (KK) tower of sterile neutrinos in supernova (SN)
physics.

The potential relevance of bulk fermions for neutrino physics has
first been pointed out and analyzed in~\cite{DDG,DS}. Fermions
propagating in large extra dimensions appear as light, dense KK towers
of sterile neutrinos in 4 dimensions, possibly mixing with the
Standard Model ones. In turn, the existence of extra dimensions much
larger than the inverse electroweak scale is natural in string
inspired models with a fundamental scale in the TeV
range~\cite{TeV,TeV3}, if the standard relation between that scale and
the Planck scale is implemented.

The basic ideas about supernova explosion and cooling, nicely
confirmed by the neutrino signal from \SN~\cite{signal}, translate
into significant constraints on the energy loss through invisible
channels~\cite{ann}. In fact, in order for the observed neutrino
signal to be accounted for, the invisible energy loss should not cool
the protoneutron star on a time scale shorter than about 10 seconds.
The bulk of a large extra dimension provides at least two interesting
examples of invisible channel: KK graviton emission~\cite{graviton}
and mixing of Standard Model and bulk neutrinos, the possibility we
are interested in. With respect to the case of a single sterile
neutrino~\cite{sterile}, conversion into KK towers of sterile
neutrinos is enhanced by the higher number of available states and
especially by matter effects~\cite{BCS}.
The constraints that follow on the parameter space of bulk neutrino
models look at first sight quite severe~\cite{BCS}. However, those
constraints are relaxed by an interesting feedback
mechanism\footnote{A different type of feedback was considered
  in~\cite{Valle}.} that prevents unacceptable energy
loss~\cite{LRRR2} and affects the protoneutron star deleptonization
and cooling through a non trivial interplay with diffusion. By
suppressing the neutrino MSW potential and then making it negative
everywhere, the mechanism leads to a self-limiting of the potentially
dangerous, MSW-enhanced, energy loss into the bulk and requires a
revisitation of the bounds in the literature.  The consequences for SN
physics, in particular explosion, deleptonization, cooling and
spectra, are not less interesting. From this point of view what we
discuss here is an example of how the standard ideas about the
protoneutron star evolution can be affected in presence of new physics
in an unexpected and intriguing way.

The paper is structured as follows. In Section~\ref{sec:framework} we
shortly review the salient points about (electron) neutrino
oscillations into bulk neutrinos. Section~\ref{sec:main} is the
central part of the paper. Since a reliable study requires taking into
account diffusion, we first set up in Section~\ref{sec:assumptions} a
toy model of the protoneutron star core which, although minimal,
contains all the neutrino transport physics relevant to our
discussion. In Section~\ref{sec:loss} we incorporate in the model the
effect of electron neutrino and antineutrino conversion into bulk
neutrinos. The effect on the evolution of the protoneutron star core
is described in Section~\ref{sec:evolution}. The consistency with \SN\ 
is shown in Section~\ref{sec:implications}, where the implications for
the neutrino luminosity, the flux spectrum and the shock reheating are
also discussed.  Section~\ref{sec:subleading} summarizes the
constraints on subleading non self-limiting conversion probabilities,
that provide one of the relevant bounds on our parameter space. We
conclude in Section~\ref{sec:summary}.

\section{Framework}
\label{sec:framework}

\subsection{The supernova}

Before embarking in the detailed analysis, let us shortly and
qualitatively review the standard ideas about SN explosion
and cooling~\cite{reviews,reviews2}.

Type II Supernovae originate from evolved, massive stars whose iron
core reaches the Chandrasekhar limit and collapses under gravitational
pressure. During the collapse, neutrinos start to be produced by
electron capture and get trapped when densities of order
$10^{12}\gcm$ are reached. The inner part of the iron core has an
homologous collapse that stops when, after a fraction of a second,
stiff nuclear densities of order $\rho_0=3\cdot 10^{14}\gcm$ are
reached. The free falling outer material then builds up a shock that
propagates outwards. It is believed that this ``prompt'' shock has not
enough energy to give rise to a successful explosion and hence stalls
after $\sim 0.1\sec$ at a radius of order $100\km$. Below the shock,
the protoneutron star has a high density inner part, the inner core,
with a radius of 10--30$\km$ and a bloated outer part, the mantle,
that accretes mass falling through the shock. Let us leave for a
moment the shock to its uncertain fate and follow the evolution of the
protoneutron star below it. Most of the gravitational binding energy
of the progenitor star, a few $10^{53}\erg$, is stored there. Only
$\ord{1\%}$ out of that will go into explosion kinetic energy or
electromagnetic radiation. Most of it will be released through
neutrino emission in a few tens of seconds.

The stiff inner core has a mass of order of the solar mass,
$\Msun\simeq 2\cdot 10^{33}\,\mathrm{g}$, a density of order $\rho_0$,
a temperature of 10--30$\MeV$ and a lepton fraction per baryon $\Y\sim
0.35$. Electrons and electron neutrinos are highly degenerate in such
conditions, with Fermi momenta of order 300 and 200$\MeV$
respectively. The mean free path of a $200\MeV$ neutrino is
$\lambda\sim 10\cm$. This determines the time scale for neutrino
diffusion and emission: $\td\sim 3 R^2/\lambda \sim 10\sec$ for a
neutrino at a depth $R=10\km$ in the inner core. Analogously, the
total neutrino flux and spectrum can be roughly estimated from the
knowledge of the star structure and temperature where neutrinos
experience their last energy exchange.  The mantle looses most of its
lepton number, accretes most of its mass, cools and contracts in only
0.5--1$\sec$. Then, as it cools, the whole protoneutron
star slowly contracts to a radius of order $10\km$.  After a few tens
of seconds, we are left with a proper neutron star.

Since supernovae do explode, we can be confident that the shock has in
the meanwhile successfully expelled the outer matter. An important
contribution to the revival of the stalled shock is given by the
energy deposition of the neutrino flow. A $\ord{1\%}$ fraction of the
energy flowing might be enough to give rise to a sufficiently
energetic explosion a second or so after collapse. Critical are of
course the efficiency of the energy transfer mechanism and the
intensity of the neutrino source.

Although this ``delayed shock'' mechanism is currently believed to
play an important role, reproducing the explosion in computer
simulations it is still non trivial at present~\cite{explosion}. This
does not necessarily mean that something is missing in our general
understanding. The problem is likely to originate from the complexity
of the system to be modeled and the variety of the physics involved.
It may just be necessary to turn to full 3D simulations able to
account for non-spherically symmetric effects. Convection is likely to
be important in this context, since it may boost the neutrino
luminosity~\footnote{However, whether such effects do provide a
  solution of the SN problem is still an open
  issue~\cite{convection}.}.  Moreover, the neutrino energy deposition
in what was the prompt shock is likely to be just one of the several
effects giving rise to the explosion~\cite{explmech} and the outcome
critically depends on a number of poorly known variables like the
progenitor structure, the equation of state, the details of neutrino
transport, etc.  We also note that a purely neutrino driven explosion
is not obviously reconciliable with the indications of a non-spherical
mechanism \cite{asymmech}~\footnote{The invisible channel we are
  discussing might itself provide a source of asymmetry.}.  Still, it
is worth pointing out that the new physics and mechanism we discuss in
this paper might also play an important role by enhancing the neutrino
luminosity. This will be discussed in Section~\ref{sec:implications}.

The basic ideas about protoneutron star deleptonization and cooling
have been nicely confirmed by the observation of a neutrino signal in
the Superkamiokande and IMB detectors in coincidence with
\SN~\cite{signal}. Despite the low statistics, there is definitely a
significant agreement between the measured neutrino flux duration and
intensity and the estimates above. This allows to set limits on
drastic departures from the described picture, in particular on
invisible energy loss channels that would waste the available energy
on timescales shorter than $10\sec$.

\subsection{Bulk neutrinos and feedback}

We consider a Kaluza-Klein (KK) tower of sterile neutrinos living in
the bulk of a single, large, flat, gravitational extra dimension.
Additional smaller dimensions will be in general present. What we
consider is therefore an effective theory valid up to scales below the
inverse scales of the smaller dimensions. The physics we are
interested in involves bulk neutrinos with masses within $\sim
100\keV$. We therefore assume that the inverse scale of the additional
smaller dimension is larger than $100\keV$ and neglect them in the
following.

As for the size $R$ of the extra dimension we consider, we only
require it to be in the range
\begin{equation}
  \label{eq:Rrange}
  10^{-3}\eV\lesssim\frac{1}{R}\lesssim 1\keV \, .
\end{equation}
The lower bound comes from tests of the Newton's law at small
distances\footnote{Depending on the specific model and on the scale of
  the mixing term $m$ in \eq{mixing} below, it may be necessary to
  strengthen the lower bound to $1/R\gtrsim 10^{-1}\eV$ in order to
  avoid undesired effects in solar and atmospheric neutrino
  experiments. In fact, for $1/R\lesssim 10^{-1}\eV$, the first modes
  of the KK tower may participate to oscillations if their masses are
  given by $n/R$, $n=0,1,2,\ldots$. This is not the case in presence
  of a Dirac or Majorana bulk mass $\mu$, since in this case all bulk
  neutrinos are heavier than $\mu$~\cite{LRRR,LRRR2}. Values of $1/R$
  as low as $10^{-3}\eV$ would then be automatically compatible with
  not having effects in the observed oscillation signals.  Provided
  that $\mu \ll\keV$, such a mass term would not affect SN physics.}
and the origin of the upper bound, as well as the bounds below, will become clear by the end of the
section.

Apart from $R$, which gives their density of states, all we need to
know about bulk neutrinos is their mixing with the electron neutrino.
We assume that the mixing $\theta_k$ of $\nu_e$ with the (almost)
sterile $k$-th mass eigenstate $N_k$ is inversely proportional to its
mass $M_k$,
\begin{equation}
  \label{eq:mixing}
  \theta_k \simeq \frac{m}{\sqrt{2}M_k}\,,
\end{equation}
a generic feature of most models in the
literature~\cite{DDG,DS,BCS,LRRR2,LRRR,list}.
\Eq{mixing} defines the mixing parameter $m$.\footnote{The exact
  dependence of $M_k$ on $1/R$ may depend on the specific model. We
  only assume the density of states $dk/dM$ to be $R$ at scales around
  $10\keV$. The parameter $m$ usually originates from Yukawa
  interactions involving the 5D neutrino, the lepton doublet and an
  Higgs field.}

We are interested in a range for $m$ such that $mR\ll 1$,
which ensures that all mixings $\theta_k$ are small, and
\begin{equation}
  \label{eq:mmRrange}
  10^{-12}\eV\lesssim \mmR\lesssim 10^{-4}\eV \,.
\end{equation}
The parameter $\mmR$ sets the scale of all the effects we discuss.
The upper bound ensures the smallness of the transition probabilities
over a mean free path and values of $\mmR$ smaller than the lower
bound give unobservable effects. The range in \eq{mmRrange} is further
restricted by the expectation that subleading contributions to the
oscillation probability give rise to the bound $mR\lesssim 10^{-5}$,
as we will discuss in Section~\ref{sec:subleading}. Together with the
upper bound on $1/R$, the latter gives $\mmR\lesssim 10^{-8}\eV$. We
will therefore concentrate on the first four orders of magnitude of
the range~(\ref{eq:mmRrange}).

For appropriate sizes of the volume associated to the smaller
dimensions, all values of $1/R$, $\mmR$ in the ranges above are
consistent with the relation $(\MP/\Mf)^2 = \prod_i 2\pi R_i\Mf$
between the volume of the extra space with the 4D Planck mass, for
appropriate sizes of the smaller dimensions.  At the same time, those
values are compatible with cosmological and astrophysical bounds from
KK graviton production before Big-Bang nucleosyntesis~\cite{graviton2}
and in supernovae~\cite{graviton}.

\myskip
 
Having described the neutrino parameter space, we now discuss neutrino
oscillations in the inner core of the protoneutron star.  Oscillations
are dominated by matter effects. While bulk neutrinos are sterile and
therefore unaffected by matter, the electron neutrinos acquire an
effective squared mass $\meff^2 = 2 E V$, where
\begin{equation}
  \label{eq:V}
  V = \sqrt{2}\GF\nB\left(\frac{3}{2}Y_e+2Y_{\nu_e}-\frac{1}{2}\right)
\end{equation}
is the MSW matter potential, $\nB$ is the baryon number density and
$Y_{e, \nu_e} = n_{e, \nu_e}/\nB$ is the number fraction of $e$,
$\nu_e$ per baryon. More precisely, $n_{e,\nu_e} =
N_{e^-,\nu_e}-N_{e^+,\bar\nu_e}$ denotes the difference of the
particle and antiparticle number densities. Also, we denote by $\Y$
the total lepton fraction $Y_e+Y_{\nu_e}$.

\begin{figure}
\begin{center}
\epsfig{file=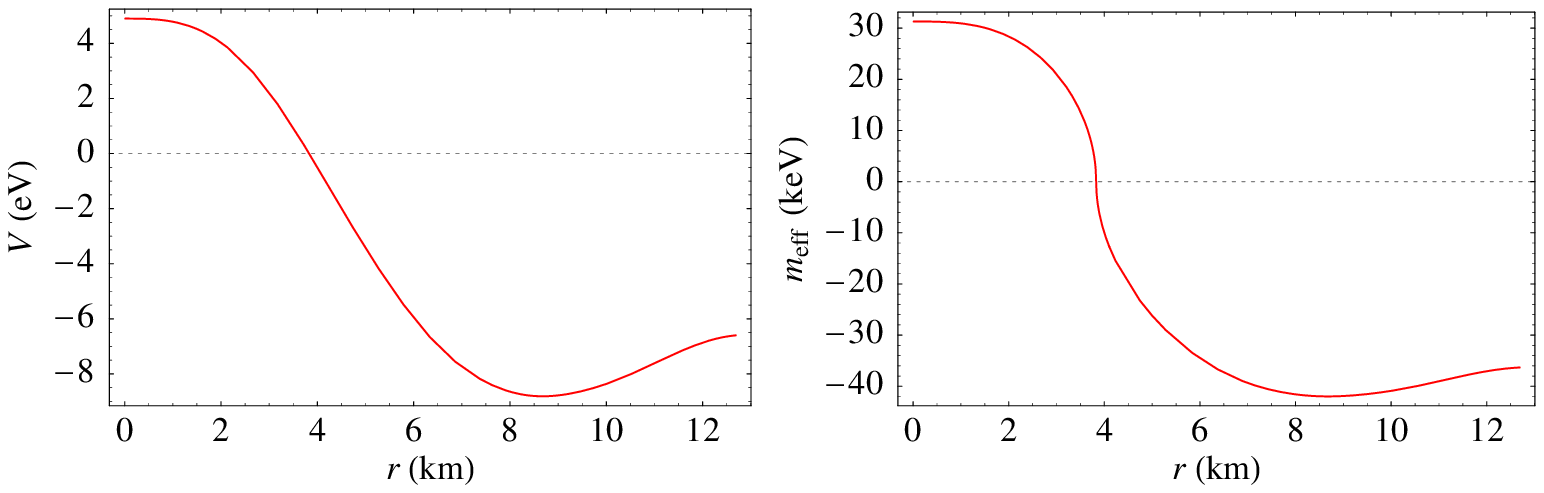,width=1.00\textwidth}
\end{center}
\begin{minipage}[t]{0.475\textwidth}
  \mycaption{Profile of the MSW potential of electron
    neutrinos corresponding to the initial profiles in \fig{fast}.}
\label{fig:V} 
\end{minipage}\hspace*{0.05\textwidth}
\begin{minipage}[t]{0.475\textwidth}
\mycaption{Effective mass of a $100\MeV$ neutrino (where $V>0$) or
  antineutrino (where $V<0$) induced by the potential in \fig{V}.}
\label{fig:meff} 
\end{minipage}
\end{figure}

The neutrino contribution to $V$ is important in the inner core. Also
important is the negative contribution from neutrons, that allows the
potential to have both signs, as shown in \Fig{V}. According to the
sign of the potential, matter effects will enhance neutrino or
antineutrino disappearance in the bulk.

The MSW potential ranges between $-10\eV$ and $5\eV$.
Correspondingly, in the early stages of evolution, most electron
neutrinos have an effective mass in the range
\[
(10\keV)^2\frac{E}{100\MeV}\lesssim \meff^2 \lesssim
(50\keV)^2\frac{E}{100\MeV}\,,
\]
where $E$ is the neutrino energy, as shown in \fig{meff}. A resonance
is met when the effective mass matches the mass of a state in the KK
tower.  The condition $mR\ll 1$ ensures that the width of the
resonance is small compared to the distance between levels.

When an electron neutrino emerges from an interaction in the inner
core, its mixing with the massive states in the KK tower depends on
its effective mass $\meff$ and therefore on the values of $\nB$,
$Y_e$, $Y_{\nu_e}$ where the interaction took place. If $V$ is
positive, $\nu_e$ is predominantly made of the mass eigenstate $N_k$
such that $M_k < \meff < M_{k+1}$, with only a small component from
the other eigenstates\footnote{This is always true except when $\meff$
  coincides with some $M_k$ within a resonance width. In such a case,
  $\nu_e$ will predominantly be made by a superposition of $N_k$ and
  $N_{k+1}$.  See~\cite{LRRR2} for a detailed discussion of mixing.}.
If $V<0$, on the other hand, $\nu_e$ does not mix with the sterile
states --- the mixing takes place in the antineutrino sector in this
case. After its interaction, the neutrino travels a distance of order
of its mean free path, say about $10\cm$, before experiencing the next
interaction. Let us determine the probability $\Ploss$ that the
electron neutrino has in the meanwhile oscillated into a bulk neutrino
assuming that $V$ is positive. If $V$ is negative, the probability is
negligible and what below applies to antineutrinos. Over $10\cm$, the
values of $\nB$, $Y_e$, $Y_{\nu_e}$ change only slightly, but the
corresponding change in the effective mass may be sufficient for the
neutrino to predominantly mix with a different mass eigenstate $N_h$.
The dominant contribution to the disappearance probability $\Ploss$ is
then given by the probability that $N_k$ has not turned into $N_h$.
This level ``non-crossing'' probability can be computed by means of
the Landau-Zener formula~\cite{DS}
\begin{equation}
  \label{eq:LZ}
  \Ploss\simeq1-e^{-\frac{\pi}{2} |\Delta n| \gamma}\simeq\frac{\pi}{2}
  |\Delta n| \gamma\,, 
\end{equation}
where $\Delta n$ is the number of levels crossed, $\gamma$ is the
adiabaticity parameter and we assumed $\Ploss\ll 1$, as follows from
$\mmR\ll 10^{-4}\eV$ (for $\Ploss\sim 1$, the analysis is complicated
by non linear effects and back conversion of bulk neutrinos into
$\nu_e$).
The Landau-Zener formula can be extended to the case of multiple
resonances because the latter are well separated.

The adiabaticity parameter is given by
\begin{equation}
  \label{eq:gamma}
  \gamma = \frac{m^2 V}{E(\partial V/\partial r)\cos\phi}\,,
\end{equation}
where $\phi$ is the angle between the neutrino momentum and the radial
direction. The number of levels crossed over a distance $L$
is\footnote{For $1/R\gtrsim 1\eV$, not all neutrinos cross a resonance
  and $\Delta n$ should be considered as the average number of level
  crossed (see below).}
\begin{equation}
  \label{eq:dn}
\Delta n = \frac{\partial V}{\partial r}\frac{R E}{\meff}\,L\cos\phi \,.
\end{equation}
We therefore conclude that \globallabel{eq:Pnc}
\begin{align}
  \Ploss &\simeq \theta(V) L\,\frac{\pi}{2\sqrt{2}}\mmR
  \left(\frac{|V|}{E}\right)^{1/2} \mytag \\
  \Plossbar &\simeq \theta(-V) L\,\frac{\pi}{2\sqrt{2}}\mmR
  \left(\frac{|V|}{E}\right)^{1/2} \,. \mytag
\end{align}
The step function $\theta$ accounts for the fact that neutrinos
(antineutrinos) do not cross any resonance if $V<0$ ($V>0$).
Correspondingly, we have neutrino disappearance for $V>0$ and
antineutrino disappearance for $V<0$. The probabilities depend on the
neutrino parameters only through $\mmR$, to which they are
proportional. The upper bound in \eq{mmRrange} ensures that $\Ploss\ll
1$.

In obtaining the previous equation, we have approximated the electron
neutrino with the mass eigenstate of which it is predominantly made.
Additional subleading contributions to $\Ploss$ come from the small
mixing with the neglected eigenstates, that amount to a fraction of
order $mR$ of the electron neutrino. Their effect will be discussed
in Section~\ref{sec:subleading}.

Let us now discuss the motivation for restricting the range of $1/R$
to $1/R\lesssim\keV$. The number of resonances crossed in a mean free
path can be estimated from \eq{dn}. Roughly speaking, for
$1/R\lesssim\eV$ most neutrinos are likely to cross at least a
resonance in a mean free path --- hundreds in the lower part of the
allowed range. For $1/R\gtrsim\eV$, on the other hand, resonances are
met only by neutrinos with energies within $\pm\delta E$ from a set of
discrete values $E_k = M^2_k/(2 V)$ in which $\meff = M_k$. The
discrete values $E_k$ are separated by $\Delta E_k = M_k/(RV)$. When
$1/R\gtrsim\eV$, $\delta E = E L \cos\phi (\partial V/\partial r)/V$
is smaller than the separation $\Delta E_k$ and the resonances are not
crossed for arbitrary values of the energy anymore.  However, there
will always be neutrinos crossing a resonance as long as some of the
energies $E_k$ fall in the neutrino spectrum. The condition
$1/R\lesssim 1\keV$ ensures that there are at least 20--50 points in
the spectrum where neutrinos cross a resonance at the beginning of the
evolution. It also ensures that there will be a few wells in the
spectrum even when, at later stages, the product $V\cdot E$ will
decrease by a factor 1000.

\myskip

We now have all the ingredients to understand how $\nu_e$ conversion
in the inner core tends to draw the MSW potential to zero everywhere
through a feedback mechanism. Let us consider a region in the inner
core where the MSW potential is positive. \Eq{Pnc} shows that the
neutrino conversion probability is proportional to $\sqrt{V}$. On the
other hand, once a neutrino has been lost in the bulk, $V$ decreases.
This is because $V\propto 3Y_e/2+2Y_{\nu_e}-1/2$. As a consequence of
the loss, the electron and electron neutrino number densities will
readjust to restore $\beta$ equilibrium, but $3Y_e/2+2Y_{\nu_e}$ will
certainly decrease. The more neutrinos are lost in the bulk the
smaller $V$ becomes. It may happen, and we will see that it does
happen, that $V$ approaches zero before all the lepton number and
energy are lost. At this point the neutrino conversion through the MSW
enhanced probability~(\ref{eq:Pnc}a) stops. In the region where the
potential is negative, antineutrino escape would take place (at a much
slower rate) drawing again the potential toward zero. As we will see,
the possibility of reaching zero depends in this case on the local
values of the thermodynamic variables.

The effect we are dealing with is potentially quite interesting.
Besides limiting the energy loss, in fact, the condition $V=0$
represents an important constraint on the relative abundances, with
significant implications for the SN phenomenology. However, several
issues have to be addressed.  First of all, the $V=0$ condition is
spoiled by neutrino diffusion, which tends to lower $V$ below zero.
Antineutrino conversion into bulk neutrinos then tends to bring $V$
back to zero at the expenses of an additional energy loss. The actual
evolution will be determined by an interplay of diffusion and
conversion.  Moreover, it has to be proven that the mechanism is
compatible with \SN\ observation.  Does the inner core loose all its
lepton number and energy in the attempt of reaching the $V=0$
condition? Or while neutrinos try to diffuse out? How quickly is the
condition approached?  Is the escape of the rare antineutrinos
efficient enough? Qualitative arguments are not sufficient to reliably
address the issues above, not to mention attacking the problem of
implications for possible future SN observations. On the other hand, a
full, detailed simulation of the protoneutron star evolution is
clearly beyond the scope of this paper.  We therefore set up in the
next section a toy model of inner core evolution which only
incorporates the basic ingredients of neutrino diffusion but contains
all the physics relevant to our study.

\section{Inner core evolution}
\label{sec:main}

\subsection{Standard evolution: equations and assumptions}
\label{sec:assumptions}

We now discuss the equations we use to model the evolution of the
inner core. We start from the case of no mixing with bulk neutrinos.
We concentrate on the inner core because the bulk of energy and lepton
number is stored there.  Moreover, the loss rates and the
effectiveness of the feedback on the MSW potential are highest in the
core. That is because the neutrino densities are large and because the
transition probability in \eqs{Pnc} grows with the potential, which is
proportional to the baryon density\footnote{The transition probability
  goes with $(V/E)^{1/2}$, where $E$ also grows with the baryon
  density $\nB$.  However, $V\propto\nB$, whereas
  $E\propto\nB^{1/3}$.}.

Focusing on the inner core has also practical advantages. Unlike the
mantle, the core settles into local thermodynamic equilibrium very
quickly after the collapse. The hydrodynamics is also much simpler,
with the mass profile becoming essentially constant in a few hundreds
milliseconds~\cite{reviews2}. The mantle of the protoneutron, on the
other hand, accretes matter for the first $0.5$--$1\sec$ and then
slowly contracts.

Assuming thermodynamic equilibrium and neglecting general relativity
effects, the basic equations of neutrino transport in absence of
mixing are~\cite{BL}
\globallabel{eq:transport}
\begin{align}
  \nB\frac{\partial\Y}{\partial t} &= -\vec\nabla\cdot\vec F_L \mytag \\
  \nB T\frac{\partial s}{\partial t} &= -\vec\nabla\cdot\vec
  F_\epsilon +\mun \vec\nabla\cdot \vec F_L \mytag \,,
\end{align}
where $s$ is the entropy per baryon and $\vec F_L = \vec F_\nu -\vec
F_{\bar\nu}$, $\vec F_{\epsilon_L} = \vec F_{\epsilon_\nu} +\vec
F_{\epsilon_{\bar\nu}}$ are the lepton number and lepton energy
density currents respectively. The neutrino currents are
\globallabel{eq:currents}
\begin{align}
  \vec F_\nu &= -\frac{1}{3}
  \sum_{\nu=\nu_e,\nu_\mu,\nu_\tau}\int\frac{d\vec p}{(2\pi)^3}
  \lambda_\nu(E)\vec\nabla f_\nu(E) \mytag \\
  \vec F_{\epsilon_\nu} &= -\frac{1}{3}
  \sum_{\nu=\nu_e,\nu_\mu,\nu_\tau} \int\frac{d\vec
    p}{(2\pi)^3}\,E\,\lambda_\nu(E)\vec\nabla f_\nu(E) \,, \mytag
\end{align}
where $\vec p$ is the neutrino momentum, $E=p$ is the neutrino energy,
$\lambda_\nu$ is the neutrino mean free path (mfp), which also depends
on the local thermodynamic variables, and $f_\nu$ is the Fermi-Dirac
distribution, $f_\nu(E) =(e^{(E-\mun)/T}+1)^{-1}$, which depends on
the temperature $T$ and the neutrino chemical potential $\mun$.
Analogous expressions hold for antineutrinos with $\mun\to-\mun$. We
assume that muon and tau neutrinos have vanishing chemical
potential\footnote{Given the high energies involved, a non-vanishing
  number of muons could participate to the inner core life, giving a
  non-vanishing chemical potential.}.  As a consequence, they do not
contribute to the lepton number current and only give a thermal
contribution to the energy current.  The neutrino chemical potential,
as the electron, neutron and proton ones, $\mu_e$, $\mu_n$ and
$\mu_p$, can be obtained in terms of the thermodynamic variables $T$,
$\Y$ and $\rho$ ($\rho$ is the mass density), by solving the
equilibrium equation $\mu_e - \mun = \mu_n - \mu_p$. Since we do not
solve the whole protoneutron star evolution, we also have to specify
boundary conditions for \eqs{transport}. We approximate them by
imposing that the lepton number and energy fluxes at the border are
proportional to the neutrino number and energy density.  We assume
spherical symmetry.

The transport equations~(\ref{eq:transport}) for $\Y$ and $T$
incorporate all the physics we are interested in before inclusion of
non-standard effects. Their solution however requires i) the knowledge
of the mean free paths, ii) the knowledge of the equation of state,
which enters the equilibrium condition and the entropy, and iii) the
determination of the density $\rho(r,t)$.  In order to prune our
considerations from unnecessary complications and concentrate on the
transport physics of \eqs{transport}, we address the three points above
in the simplest possible way. First of all, we take advantage of the
hydrodynamic stability of the inner core by using a constant (in time)
density profile $\rho(r)$.  As for the equation of state, we consider
a core made only of $n$, $p$, $e^\pm$, $\nu_{e,\mu,\tau}$,
$\bar\nu_{e,\mu,\tau}$, $\gamma$ and we take into account the strong
nuclear interactions through a modified dispersion relation for the
nucleons. That is, the nucleons are approximated as a perfect Fermi
gas, with an effective mass $m_N^*$ carrying a dependence on the state
variables.  A detailed calculation of $m_N^*$ has to rely on some
effective theory of nucleon-nucleon interaction and, in general, will
deviate substantially from the vacuum value $m_N = ~939 \MeV$.  The
result of such theories can be reproduced by a simple expression
carrying only a dependence on the medium density~\cite{reviews}
\begin{equation}
  \label{effmass}
  m_N^* = \frac{m_N}{1+\beta_0 ~\rho/\rho_0}
\end{equation}   
where $\beta_0$ is chosen to be 0.5 and $\rho_0 = 3 ~10^{14}$
gr/cm$^3$ is the reference nuclear density.

Finally, we have to define a prescription for the mean free path of
neutrinos of different flavors.  For electron neutrinos the most
important contributions to the opacity come from the absorption
reaction $\nu_e + n \rightarrow e^- + p$ and from the scattering on
neutrons.  Scattering on protons and, to a smaller extent, on
electrons, are also sizable.  For $\nu_\mu$'s and $\nu_\tau$'s, only
the scattering reactions are important.  A detailed analysis of
neutrino opacities should take into proper account the cross sections
for these and several other subleading processes, as well as the
status of the nuclear matter (degenerate or non-degenerate nucleons,
consequent Fermi blocking effects), the nucleon-nucleon interactions,
the multicomponent nature of dense matter (e.g. the presence of
hyperons), etc.  For a comprehensive discussion on this subject we
refer the reader to~\cite{Reddy} and references therein.  For our
purposes, it is sufficient to model the neutrino mean free path with a
simple inverse quadratic dependence on the neutrino energy
\begin{equation}
  \label{eq:mfp}
  \lambda_{\nu_e}(E)=\lambda_{\nu_e}^0 \frac{E_0^2}{E^2}\,, 
  \qquad  \lambda_{\nu_\mu,\nu_\tau}(E)
  =\lambda_{\nu_\mu,\nu_\tau}^0 \frac{E_0^2}{E^2}\,,
\end{equation}
where we take $\lambda_{\nu_e}^0 = 1.2$ cm and
$\lambda^0_{\nu_\mu,\nu_\tau} = 4.8$ cm at the reference energy $E_0 =
260 \MeV$.  The values of $\lambda_{\nu_e}^0$ and
$\lambda^0_{\nu_\mu,\nu_\tau}$ determine the diffusion time scale.
When compared with the exact general expression of ref.~\cite{Reddy}
(taken at typical density, temperature and leptonic fraction), the
above proves to be a good approximation for the upper portion of the
range of neutrinos' energies of our interest while it turns out to be
poorer in the lower part.  Moreover, the above choices yield an
evolution whose timescale agrees with the results of more
sophisticated analysis~\cite{Pons}.

With the choice above on the energy dependence of the mean free paths
and setting $\lambda_{\bar\nu}=\lambda_\nu$ in the small antineutrino
contribution, one finds the simple expressions
\begin{equation}
  \label{eq:currents2}
  -\vec F_L = a_e \vec\nabla\mun\,, \qquad
  -\vec F_{\epsilon_L} = \frac{a_e}{2}\vec\nabla\mun^2 +(a_e+a_\mu+a_\tau)
    \frac{\pi^2}{6}\vec\nabla T^2 \,, 
\end{equation}
where $a_i = \lambda^0_{\nu_i} E_0^2/(6\pi^2)$, $i= e, \mu,\tau$.

As mentioned, the simplified model we use for discussing the inner core evolution only contains the basic features of neutrino diffusion relevant to our discussion.
It is certainly not suitable for addressing more complex issues such as non-spherically symmetric effects likely to be involved in the explosion.

\subsection{Neutrino escape into the bulk}
\label{sec:loss}

Eqs.~(\ref{eq:transport},\ref{eq:currents2}), together with the
equilibrium equations and the boundary conditions represent our
simplified model of inner core deleptonization and cooling in absence
of non standard dynamics.  We now incorporate the effect of neutrino
and antineutrino escape in the bulk.

Electron neutrinos with energy $E$ contribute to the lepton number
loss rate by $\vev{P}/\lambda$ each, where $P=\Ploss$ and ``$\vev{}$''
indicates an average over the distance traveled $L$.\footnote{Note the
  difference between $\vev{P}/\lambda$ and $\vev{P/L}$. While the
  latter expression looks at first sight more correct, it actually
  does not take into account, as the former does, the variation of the
  number densities in the time necessary to travel the distance $L$.}
Since in our case the probability is linear in $L$, we simply have
$\vev{P}/\lambda = P/L = \theta(V)\pi/(2\sqrt{2})\mmR (|V|/E)^{1/2}$.
The total neutrino number loss rate $\Gamma_\nu$ follows from
integration over neutrino momenta and can be written by means of the
Fermi integrals $F_\alpha$ as
\begin{gather}
  \label{eq:nloss}
  \Gamma_\nu = \frac{\theta(V)}{4\sqrt{2}\pi}\mmR\sqrt{|V|} T^{5/2}
  F_{3/2}\left(\frac{\mun}{T}\right)\,,\quad \Gamma_{\epsilon_\nu} =
  \frac{\theta(V)}{4\sqrt{2}\pi}\mmR\sqrt{|V|} T^{7/2}
  F_{5/2}\left(\frac{\mun}{T}\right)\,, \notag \\
  F_\alpha(y) = \int_0^\infty dx \frac{x^\alpha}{e^{x-y}+1}\,.
\end{gather}
Analogously for antineutrinos.  Putting all together, we find the
following modified evolution equations for $\Y$ and $T$:
\globallabel{eq:evolution}
\begin{align}
  \nB\frac{\partial\Y}{\partial t} &= \vec\nabla\cdot(a_e~\vec\nabla\mun) -
  \frac{\sigma}{4\sqrt{2}\pi} \mmR\sqrt{|V|}~T^{5/2}
  F_{3/2}\left(\sigma\frac{\mun}{T}\right) \mytag \\
  \begin{split}
    \nB T\frac{\partial s}{\partial t} &= a_e\left(\vec\nabla
      \mun\right)^2+\vec\nabla\cdot\left((a_e+a_\mu+a_\tau)\frac{\pi^2}{6}\vec\nabla T^2\right) \\
    &\quad-\frac{1}{4\sqrt{2}\pi} \mmR\sqrt{|V|}~T^{7/2}\left(
      F_{5/2}\left(\sigma\frac{\mun}{T}\right)-\sigma\frac{\mun}{T}
      F_{3/2}\left(\sigma\frac{\mun}{T}\right)\right)\,,
  \end{split} \mytag
\end{align}
where $\sigma = \sign(V)$. We recognize in eq.~(\ref{eq:evolution}b)
the heating term $a_e(\vec\nabla\mun)^2$ associated to the degradation
of the degeneracy energy of the neutrinos reaching regions with lower
chemical potential. This term partially originates from the chemical
potential contribution to $T\,ds$. With the same origin, there is a
heating term in the energy loss part proportional to $F_{3/2}$ that
can overcome the cooling term proportional to $F_{5/2}$. In this case
the temperature increases because of the degradation of the degeneracy
energy of neutrinos at the Fermi surface that downscatter to replace
neutrinos inside the Fermi sphere escaped in the bulk.

In \eqs{evolution} the potential $V$, as $\mun$ and $s$, is a function
of the thermodynamic variables $\rho$, $T$, $\Y$. A simple
approximation is useful to understand the dependence of $V$ on those
variables, in particular $\Y$. By neglecting the $\mu_n-\mu_p$ term in
the beta equilibrium equation $\mu_e - \mun = \mu_n - \mu_p$, one
obtains $Y_e = 2/3\,\Y$, $Y_\nu = 1/3\,\Y$. Such a rough approximation
turns out to be pretty good for $V$, where errors cancel to give, with
an accuracy of 5\% or less,
\begin{equation}
  \label{eq:V0}
  V\simeq \sqrt{2}\GF\nB\left(\frac{5}{3}\Y-\frac{1}{2}\right) \,,
\end{equation}
which vanishes for $\Ya\simeq 3/10$ independently of $T$, $\rho$.
\Eq{V0} illustrates what happens to the MSW potential when a neutrino
or an antineutrino is lost in the unit volume. The lepton number lost
or gained is redistributed by the beta equilibrium between neutrinos
and electrons and the net effect is a variation of the potential by
the amount $\mp (5/3)\sqrt{2}\GF$. Moreover, through \eq{V0} it is
possible to explicitly see the effect of the feedback. Let us consider
the limit in which the loss term dominates the evolution and diffusion
is negligible. Then, when $\Y>3/10$, the RHS is negative and $\Y$
decreases. When $\Y<3/10$ the RHS is positive and $\Y$ increases.
When $\Y=3/10$, the RHS vanishes and the lepton fraction is constant.
In the approximation above, $3/10$ is the value of $\Y$ at which the
non standard terms in \eqs{evolution} aim. We will see that for
$\mmR\gg 10^{-12}\eV$ this value is indeed reached and maintained for
some time in some region of the inner core. In other regions, the
value is not reached but its attraction on the lepton fraction affects
the deleptonization. In our numerical implementation of
\eqs{evolution}, we do not make use of the
approximation~(\ref{eq:V0}). The $V=0$ condition then becomes a
constraint on $\rho$, $T$, $\Y$ that translates into $\Y =
\Ya(\rho,T)$.

\subsection{Phases of evolution}
\label{sec:evolution}

\begin{figure}
\begin{center}
\epsfig{file=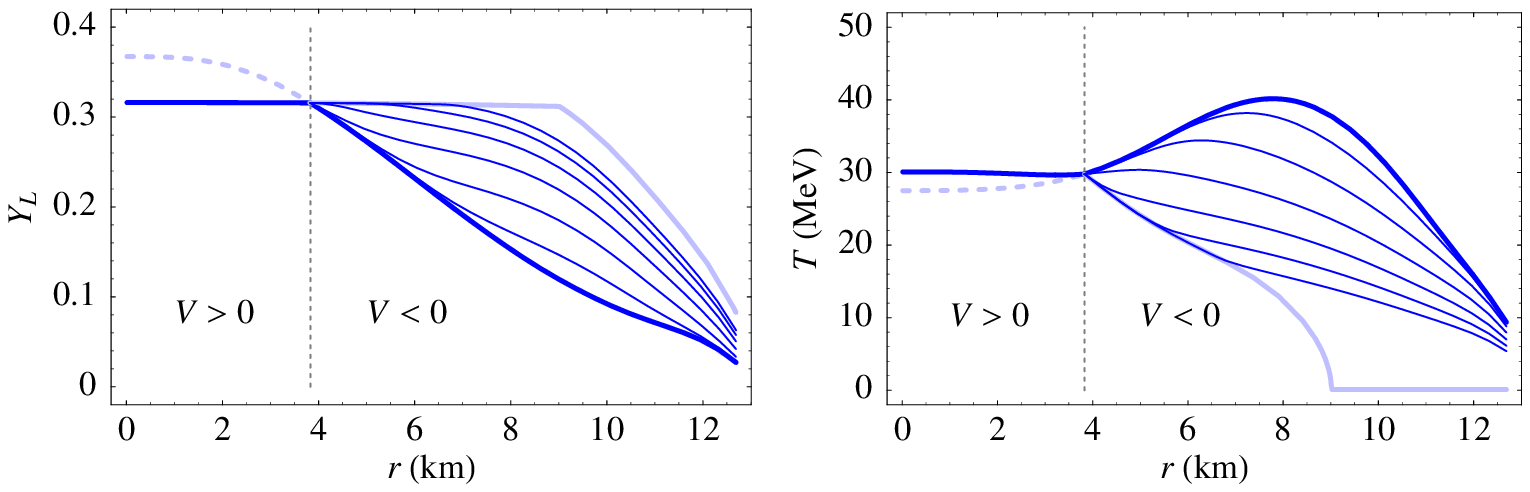,width=1.00\textwidth}
\end{center}
\mycaption{Lepton fraction and temperature profiles in the initial
  phase of the evolution. The thick dashed lines (dark solid in the
  $V<0$ region) represent the initial profiles. The dark thick lines
  show the profiles at the end of the neutrino escape phase.  The thin
  lines show the effect of antineutrino escape 1 second later for
  $\mmR = 10^{-(10,\,9,\,8,\,7,\,6,\,5)}\eV$. They can also be
  considered as the profiles at times
  $t=10^{-(5,\,4,\,3,\,2,\,1,\,0)}\sec\,(\mmR/(10^{-5}\eV))$ provided
  that $t \lesssim 1\sec$.  Finally, the light thick lines show the
  values that would be asymptotically approached if diffusion could be
  indefinitely neglected.}
\label{fig:fast} 
\end{figure}

We start to follow the evolution of the inner core a few hundred
milliseconds after collapse.  We consider a core of $1.5\,\Msun$ with
a radius of $12.7\km$ and a typical density profile described by
$\rho(r)=\rho_c/(1+(r/\bar{r})^3)$, with $\rho_c=7.5 ~10^{14}$
gr/cm$^3$, $4/3 ~\pi ~\bar{r}^3 ~\rho_c = 1.1\,\Msun$. The initial
profiles for $T$ and $\Y$ are shown in \fig{fast} (the thick dashed
lines). Their main features follow from models of core
collapse~\cite{Pons}. The detailed structure of those profiles is not
important for our purposes. Diffusion will in fact soon smooth them.
Moreover, the mixing with bulk neutrinos, when important, also affects
the early stages of SN evolution. This point will be discussed in more
detail below.

Three processes contribute to the variation of the lepton fraction:
diffusion, neutrino and antineutrino conversion into bulk neutrinos.
Each of them has a characteristic time scale, $\td$, $\tn$ and $\tnb$
respectively, defined e.g.\ as the elapsed time per fraction of $\Y$
variation. While $\tn$ turns out to be finite and roughly the same in
all the region where $V>0$, $\tnb$ strongly depends on the position in
the $V<0$ region and may become infinite, as we will see. In any case,
$\tn\ll\tnb$. In most of the inner core, in fact, the beta equilibrium
forces $\mun$ to be positive and the neutrinos to be degenerate,
$\mun\gg T$. As a consequence, the antineutrino number is strongly
suppressed and so is the lepton number variation due to antineutrino
escape.

While the time scale for diffusion is typically $\sim 10\sec$, the
escape time scale depends on $\mmR$.  Moreover, neglecting diffusion,
the evolution due to the escape depends on $\mmR$ only through an
overall scale factor, $\Y(r,t;k\,\mmR) = \Y(r,k\,t;\mmR)$,
$T(r,t;k\,\mmR) = T(r,k\,t;\mmR)$. The larger $\mmR$, the faster the
variation of $\Y$:\footnote{For $\mmR\gtrsim 10^{-7}$ the thermal
  processes would not be fast enough to maintain equilibrium. The time
  scale above would be affected, but the results below would not.}
\begin{equation}
  \tn \sim 10\sec\,
  \frac{10^{-12}\eV}{\mmR}\fracwithdelims{(}{)}{\rho}{\rho_0}^{1/2}
  \fracwithdelims{(}{)}{\Y}{0.3}^{1/3} \,.
\end{equation}
We therefore distinguish
two regimes:
\begin{itemize}
\item $\mmR \ll 10^{-12}\eV$, or $\tn\gg\td$. Neutrino
  and antineutrino escape are too slow to affect significantly the
  evolution of the protoneutron star. Consistency with the observed
  neutrino signal from \SN\ is ensured. 
\item $\mmR \gg 10^{-12}\eV$, or $\tn\ll\td$. Neutrino escape changes
  $\Y$ faster than diffusion. The consistency with the \SN\ signal
  would be in question if the feedback were not taken into
  account. Deleptonization and cooling are affected. 
\end{itemize}
In the remaining of this subsection, we will concentrate on the case
of fast escape, say $\mmR\gtrsim 10^{-11}\eV$.

We can distinguish two phases in the evolution. In the first one neutrino
escape dominates and diffusion can be neglected. The second is a mixed
phase where antineutrino escape dominates in some region and for some
time and is comparable to diffusion elsewhere. 

\myskip

The first phase takes place in the $V>0$ region, corresponding to
$\Y>\Ya$, on the left of the vertical lines in \fig{fast}. The change
of $\Y$ and $T$ due to antineutrino conversion (where $V<0$) and
diffusion is negligible, since the time scales are slower.  Due to
neutrino disappearance, both $\Y$ and the energy density start
decreasing. The essential point is whether the value of $\Y$ that
stops the MSW conversion is reached before all the energy has been
lost. This turns out to be the case in all the region where $V>0$. In
fact, by neglecting the diffusion terms in \eqs{evolution} we find the
following relation between lepton fraction and the entropy variation:
\begin{equation}
  \label{eq:TvsY}
  \frac{\partial s}{\partial\Y} = 
  \left(\frac{F_{5/2}(\mun/T)}{F_{3/2}(\mun/T)}-\frac{\mun}{T}\right) \sim
  -\frac{2}{7}\frac{\mun}{T} \;,
\end{equation}
where the partial derivative is taken at a fixed position and the
approximation $\partial s/\partial\Y\sim -2/7\,\mun/T$ holds for
degenerate neutrinos. Since such an approximation does hold for
$\Y\gtrsim0.3$ and $T\gtrsim 30\MeV$, \eq{TvsY} shows that the
temperature in the $V>0$ region actually grows while it deleptonizes.
In turn, that means that the $V=0$ condition is comfortably attained
spending only a fraction of the available energy.  Within a time
$t=0.5$--$1\sec (10^{-11}\eV/\mmR)$, $\Y$ has reached $\Ya$ in all the
$V>0$ region and the fastest phase of evolution has ended. The
profiles are now shown as solid thick lines in \fig{fast}.  The
temperature has increased as a consequence of the conversion of
degeneracy energy into thermal energy. As mentioned in
Section~\ref{sec:loss}, such a conversion takes place when states deep
inside the neutrino Fermi sphere are emptied by neutrinos escaping in
the bulk. In summary, after the initially dramatic energy leak into
the bulk, most of the energy in the $V>0$ region gets locked as $\Y$
reaches $\Ya$, which happens on the $\tn$ time scale. The thick lines
in \fig{fast} can therefore be considered as the initial condition for
the subsequent evolution.

Before discussing the next phase, a comment on the initial condition
we used is in order. As mentioned at the beginning of the subsection,
for $\mmR\gg 10^{-12}\eV$ the initial profiles are likely to be
affected by the neutrino escape. For $\mmR = 10^{-8}\eV$, for example,
moving from the initial to the $\Y=\Ya$ profile only takes about a
millisecond. Therefore, the initial profiles in \fig{fast}, following
from analysis that do not take into account the new effect, cannot be
trusted hundreds milliseconds after core bounce.  On the other hand,
given the fixed point character of the evolution in the fast phase, we
can be confident that the thick solid profile is indeed reached as
soon as the core settles. We are therefore entitled to use that
profile in the subsequent evolution. Why not to use it in the first
place and ignore the fast phase of evolution, then? Because the
simulation based on \eq{TvsY}, although it does not take into account
the conditions met in the early stages of the SN, still gives a
conservative estimate of the temperature variation and of the amount
of the energy loss, a key piece of information for what follows. In
summary, the thick profiles in \fig{fast} can be considered realistic
initial profiles for the next phase within the approximations and
uncertainties associated to our approach and to core collapse models.
Needless to say, the original initial profiles are suitable as initial
conditions for the standard diffusive regime ($\mmR\lesssim 10^{-12}$
case) within the same uncertainties.

\myskip

After neutrinos in what was initially the $V>0$ region have been
locked, antineutrinos loss begins to be significant in the $V<0$
region. This happens on the slower time scale $\tnb$, which depends on
the position in the core. Since antineutrinos are lost, the lepton
fraction $\Y$, initially smaller than $\Ya$, grows and $V$ becomes
less and less negative. Again, the essential point is whether the
$V=0$ condition, or $\Y=\Ya$, is reached before all the available
energy is lost. The relevant equation is in this case
\begin{equation}
  \label{eq:TvsY2}
  \frac{\partial s}{\partial\Y} = 
  -\left(\frac{F_{5/2}(-\mun/T)}{F_{3/2}(-\mun/T)}+
  \frac{\mun}{T}\right) \sim -\left(\frac{\mun}{T}+\frac{5}{2}\right)
  \;. 
\end{equation}
The loss turns out to be more pronounced than in the previous phase.
Actually, the energy loss per unit of lepton number change is smaller,
since the average antineutrino energy is $5/2\, T$ whereas the average
neutrino energy was $5/7\mun$. However, the entropy variation is
larger since the increase of $\Y$ leads to an increase of the energy
stored in the degenerate neutrino or electron sea at the expenses of
the temperature and entropy. That is why the RHS in \eq{TvsY2} is
larger than in \eq{TvsY}. Furthermore, the change in $\Y$ needed to
reach $\Ya$ is larger in the external part of the inner core, where
$\Y$ is lower.  In order to clarify the situation, we need to
numerically solve \eq{TvsY2}.  The result is that $V=0$ is reached
before all the local energy is gone only in the inner part of the
$V<0$ region. The outer part cools completely while $\Y$ is still
lower than $\Ya$, as it is apparent from \Figs{fast}. There, the light
thick lines represent the asymptotic lepton fraction and temperature
profiles that would be reached if the antineutrino escape phase
continued indefinitely without diffusion. In the $r\gtrsim 9\km$
region, the asymptotic temperature profile reaches zero.
Correspondingly, the maximum value of $\Y$ attainable before complete
cooling is not $\Ya$ anymore.  In practice, complete cooling does not
occur. The profiles reached after one second, before diffusion becomes
essential, are shown by the thin lines in the $V<0$ regions of
\figs{fast}. The lines correspond to
$\mmR=10^{-(10,\,9,\,8,\,7,\,6,\,5)}\eV$.  The three largest values,
that are not in the parameter space we are interested in, illustrate
the asymptotic behavior of the profiles.  Alternatively, they can be
considered as the profiles at times $t=
10^{-(5,\,4,\,3,\,2,\,1,\,0)}\sec\,(\mmR/(10^{-5}\eV))$ for a given
value of $\mmR$ --- as long as $t\lesssim 1\sec$, of course. The $V=0$
condition is reached and antineutrinos are locked in a portion of the
initially $V<0$ region which is larger the larger is $\mmR$. In a
sufficiently large but finite time that would happen in all the region
where the asymptotic $\Y$ profile is constant and the $T$ one is non
vanishing.  In the outer part of the inner core, on the other hand,
neither $\Ya$ nor the asymptotic profile will ever be reached.
However, the attraction of the fixed point on $\Y$ has the effect of
raising the $\Y$ profile, with interesting implications that will be
discussed in section~\ref{sec:implications}. The total energy loss in
this phase is again a small fraction of the total available energy.

\myskip

After the first, fast phase of neutrino escape and the second slower
phase of antineutrino escape, diffusion begins to play a significant
role. This happens on the slowest time scale $\td$. Diffusion
``unlocks'' the energy stored in the $V=0$ region. Some of it gets
lost in the bulk, the rest diffuses out in the outer part and
eventually is emitted mainly as active neutrinos. The diffusion regime
however is different from the standard one that takes place in absence
of transitions into the bulk. We have in this case a mixed regime in
which the evolution is determined by an interplay of diffusion and
escape. Let us consider first the inner part of the core, where $V=0$.
Due to diffusion, the densities change, thus spoiling the $V=0$
condition. The lepton fraction decreases and $V$ becomes slightly
negative. As $V$ becomes negative, though, antineutrinos start to
escape in the bulk thus giving a positive contribution to $V$. At some
point, the positive contribution to $V$ from conversion into bulk
neutrinos will balance the negative contribution from diffusion. In
this regime, some of the energy is lost in the bulk and some is
diffused out. The amount of energy lost in the invisible channel turns
out to be independent of $\mmR$ as long as the conversion of
antineutrinos (slower than the neutrino conversion) is efficient
enough to keep $|V|$ small.  The evolution in such a regime is well
described by a single equation for the temperature. In fact, denoting
$\Y=\Ya+\dY$, we have
\begin{equation}
  \label{eq:Vexp}
  V = V(\rho,T) \simeq \frac{\partial V}{\partial\Y}\dY \;.
\end{equation}
When inserted in \eqs{evolution}, the previous expression allows to
recover $\dY(\rho,T)$ from \eqi{evolution}{a}. \Eqi{evolution}{b} then
becomes 
\begin{equation}
  \label{eq:Tevolution}
  \begin{split}
  \nB T\frac{\partial s}{\partial t} = a_e\left(\vec\nabla
  \mun\right)^2&+\vec\nabla\cdot\left((a_e+a_\mu+a_\tau)\frac{\pi^2}{6}\vec\nabla T^2\right) \\ &+ \left(
  \mun-\sigma T \frac{F_{5/2}(\sigma\mun/T)}{F_{3/2}(\sigma\mun/T)}
  \right)\left( \vec\nabla\cdot\left(a_e~\vec\nabla\mun\right)-\nB\frac{\partial\Y}{\partial
  t}\right)\; .
\end{split}
\end{equation}
\Eq{Tevolution} always holds.  In the small $|V|$ limit, one can
approximate $\Y\simeq\Ya$ everywhere, in which case the equation
becomes selfconsistent and the evolution becomes independent of
$\mmR$. This of course will not hold forever since the efficiency of
antineutrino conversion decreases as the star cools. Moreover, it
certainly does not hold in the outer region where $V$ is well below
zero in the first place.

\begin{figure}
\begin{center}
\epsfig{file=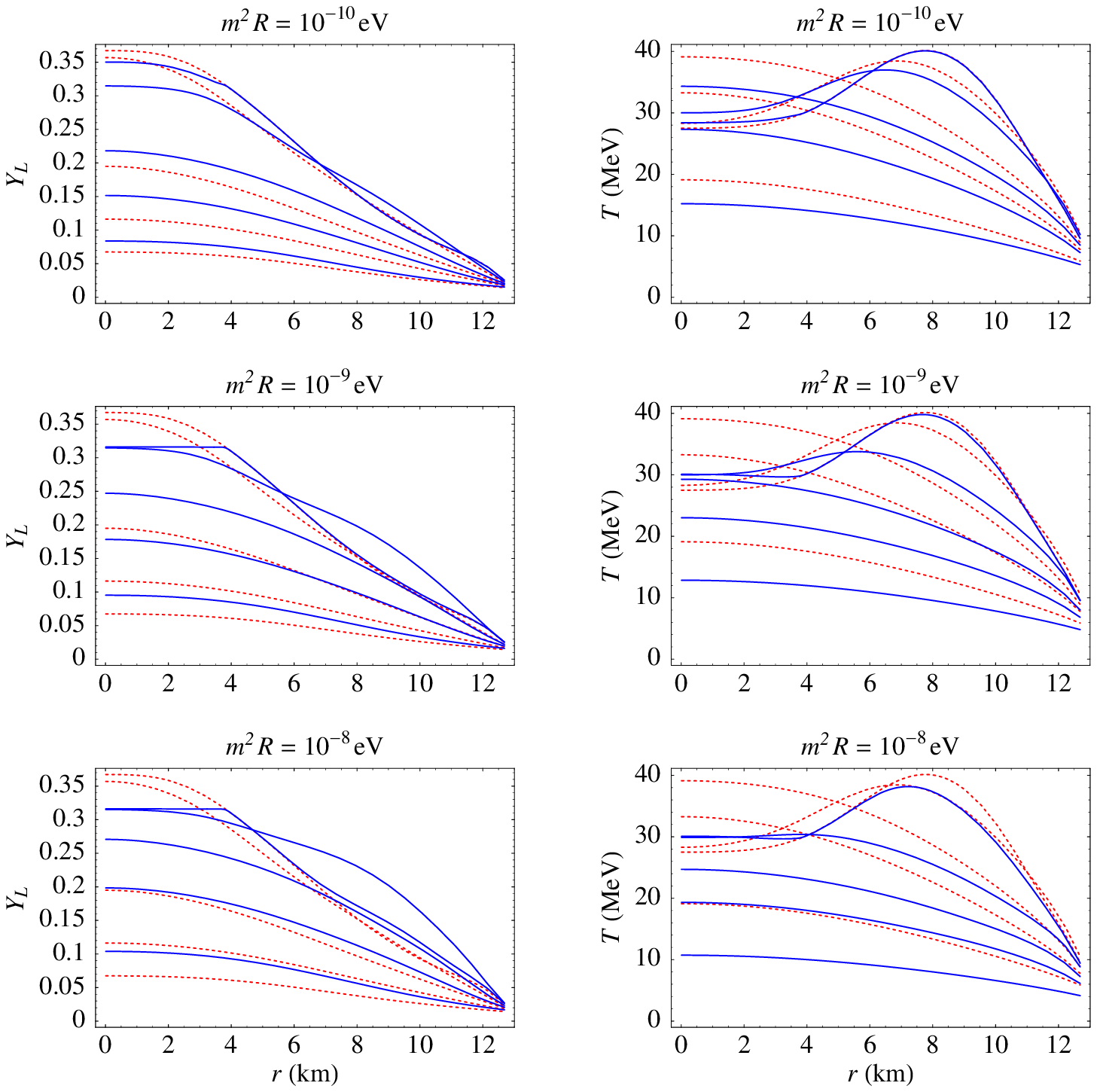,width=1.00\textwidth}
\end{center}
\mycaption{Lepton fraction and temperature profiles at
  $t=(0.01,\,1,\,10,\,20,\,40)\sec$ for different values of $\mmR$
  (solid lines). The profiles for $\mmR=0$ are also shown for
  comparison (dashed lines).}
\label{fig:evolution} 
\end{figure}

In \fig{evolution} we show the evolution of the $\Y$ and $T$ profiles
that follows from \eqs{evolution} for three values of $\mmR$:
$10^{-10}\eV$, $10^{-9}\eV$ and $10^{-8}\eV$. The case of no new
physics is also showed in each plot for comparison (dashed lines).  We
do not aim at reproducing the results of the most sophisticated
analysis in the literature for the latter case. However, a comparison
between the solid and dashed lines in figure well illustrates the
effect of new physics on the evolution. Due to neutrino conversion,
$\Y$ quickly drops to $\Ya$ in the inner part of the core.  Then, on a
longer time scale, diffusion further lowers $\Y$, thus starting
antineutrino conversion. For large values of $\mmR$, the latter is
efficient enough to keep $\Y$ close to $\Ya$ for a few seconds, until
the region becomes too cool to sustain the necessary conversion. After
the initial dramatic fall, the lepton fraction is higher than in the
case of pure diffusion. The same happens in the outer part of the
core, where the effect is more pronounced because neutrino loss never
took place and $|V|$ is larger during antineutrino conversion. The
lepton fraction even grows in the first second and then decreases
slower than in the case of pure diffusion, at the expenses of a
quicker cooling.  The consequences of this effect will be discussed in
the next subsection. For $\mmR = 10^{-8}\eV$, the temperature in the
inner core essentially never increases.

\subsection{Implications for supernova physics}
\label{sec:implications}

\begin{figure}
\begin{center}
\epsfig{file=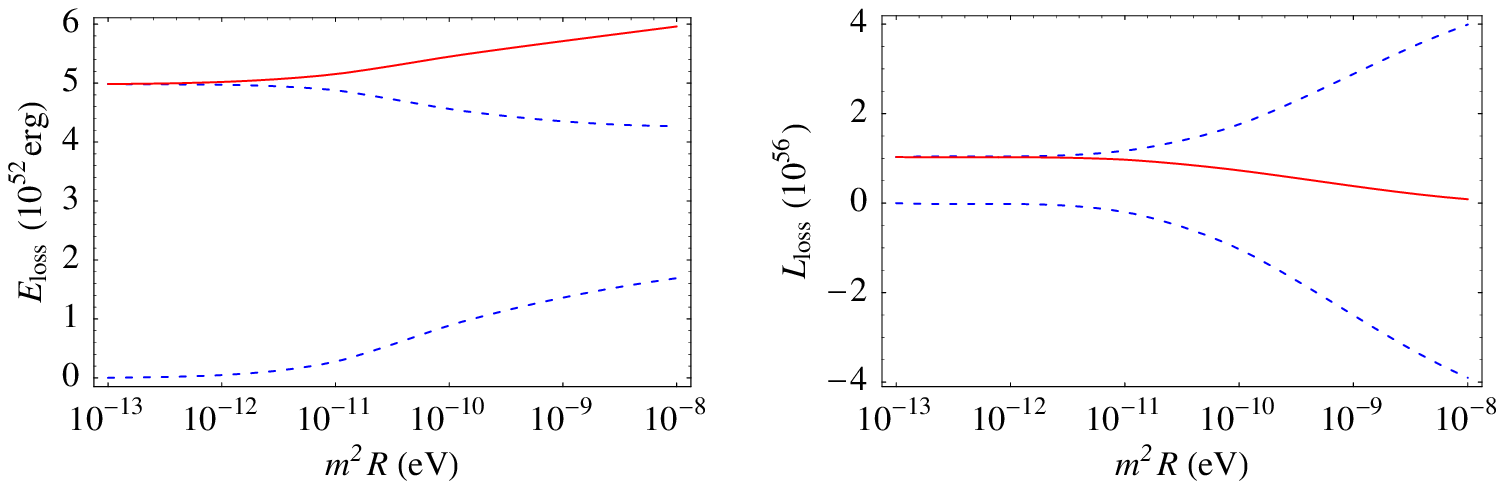,width=1.00\textwidth}
\end{center}
\mycaption{Energy and lepton number loss in the first 10 seconds. The
  total loss (solid lines) is shown together with its visible (upper
  dashed lines) and invisible (lower dashed lines) components as a
  function of $\mmR$.}
\label{fig:loss} 
\end{figure}

\Fig{evolution} shows that deleptonization and cooling do take place
on the same time scale as in absence of new physics despite the time
scale of neutrino disappearance can be orders of magnitude faster than
the diffusion one. What cannot be inferred from \fig{evolution} is the
size of the portion of the available lepton number and energy that
disappears in the bulk. Is something left to give rise to the \SN\ 
neutrino signal and to revive the shock in the delayed explosion
scenario? This issue is addressed in \fig{loss}, where the energy and
lepton number lost by the inner core in the first 10 seconds are
plotted against $\mmR$ and split in the component that goes into the
observable neutrino flux and the component that is irremediably lost
in the bulk. The latter grows with $\mmR$ but is never larger than
25\% of the total. The total energy loss also grows with $\mmR$. This
is just a consequence of the faster evolution of the temperature
profile due to the additional loss (see \fig{evolution}).  The net
effect of the growth of both the total and the invisible energy loss
is that the energy emitted in the visible channel in the first 10
seconds (as well as in the following 10 seconds) remains within 10\%
of the $\mmR = 0$ value.

\begin{figure}
\begin{center}
\epsfig{file=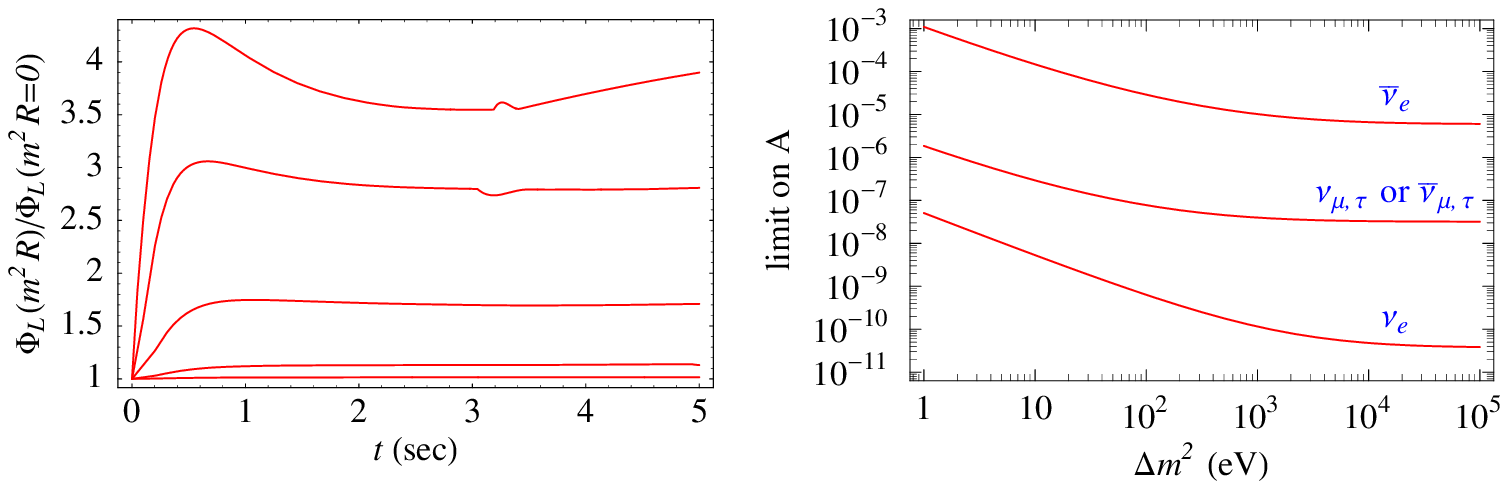,width=1.00\textwidth}
\end{center}
\begin{minipage}[t]{0.475\textwidth}
  \mycaption{Enhancement of the lepton number flux over the $\mmR=0$
    case. The five lines correspond to
    $\mmR=10^{-(12,\,11,\,10,\,9,\,8)}\eV$. The enhancement grows with
    $\mmR$.}
\label{fig:flux} 
\end{minipage}\hspace*{0.05\textwidth}
\begin{minipage}[t]{0.475\textwidth}
  \mycaption{Limits on the amplitude of $\nu_e$, $\bar\nu_e$ and
    $\nu_{\mu,\tau}$ or $\bar\nu_{\mu,\tau}$ oscillations into an
    invisible channel as functions of the squared mass difference
    $\dm{}$.}
\label{fig:limits} 
\end{minipage}
\end{figure}

The lepton number lost in the bulk in the first 10 seconds is
negative, a consequence of the dominance of the antineutrino escape
phase. The reluctance of the lepton fraction to get away from the $V =
0$ value shows in \fig{loss} in the reduction of the total lepton
number loss with $\mmR$. It is also apparent how the large negative
loss in the invisible channel translates into an enhancement up to a
factor four of the emitted lepton number. \Fig{flux} shows that the
enhancement takes place on the time scale of the initial antineutrino
loss rate and is therefore effective already in the first second.

The enhancement of the emitted lepton number has interesting
consequences on the SN phenomenology. In order to discuss them, we
first comment on the temperature of the neutrinosphere and the
spectrum of the neutrino flux. A quantitative analysis of this and the
following issues would require solving the evolution of the mantle,
which is beyond the scope of this paper. We will therefore confine
ourselves to some qualitative considerations. The disappearance
probability is negligible at the density and temperature $T_\nu$ of
the neutrinosphere, thus the latter can hardly be affected.  Moreover,
$T_\nu$ might even increase due to compressional heating until the
mantle settles. The faster inner core cooling will therefore affect
$T_\nu$ only later on. Finally, as the temperature of the
neutrinosphere decreases, the neutrino mean free path grows and the
neutrinosphere moves in the hotter interior, partially compensating
the temperature change. We do not expect the spectrum of the neutrino
flux to be significantly affected either.

The composition of the neutrino flux depends in first approximation on
$T_\nu$ and on the energy radiated per unit of lepton number.
Neglecting the difference among the neutrino
spectra~\cite{Keil:2002in}, the number of neutrinos emitted per unit
of lepton number is $N\sim k E/(LT_\nu)$, where $E$ is the energy
radiated with the lepton number $L$ and $k$ characterizes the shape of
the energy spectrum at the neutrinosphere. In absence of new physics
one has $N\sim 10$. Since the total electron lepton number of those
neutrinos must be 1, there is a slight prevalence of electron
neutrinos over the other species.  Consider now our situation in the
extreme limit $\mmR = 10^{-8}\eV$.  The lepton number emitted in say
the first second is 4--5 times larger than in the standard case, while
the energy emitted is about the same.  Each electron neutrino emitted
will be therefore accompanied in average by only a couple of neutrinos
of other species. Such a peculiar neutrino flux composition could be
indirectly observed in the existing neutrino detectors when the signal
of a SN explosion in our galaxy will reach us.

The prevalence of electron neutrinos in the neutrino flux for large
$\mmR$ has two further consequences. First, it reduces the electron
antineutrino component, which is constrained by the \SN\ signal.
Moreover, the electron neutrinos are much more effective than muon or
tau neutrinos in depositing energy in the matter outside the
neutrinosphere. As a consequence, a larger $\nu_e$ component might
help a stalling shock in ejecting the SN envelope. In fact, the
heating rate of matter irradiated by a hotter $\nu_e$ flux is
proportional to $T_\nu^2$, which we expect not to be significantly
affected by a large $\mmR$, and to the $\nu_e$ luminosity. As the left
panel of \fig{loss} shows, the total luminosity in all neutrino
species (integrated over 10 seconds, but the same holds for the
luminosity itself) is not significantly affected. Therefore, an
increase of the $\nu_e$ component has the effect of increasing the
$\nu_e$ luminosity and, in turn, the heating rate. The enhancement can
be significant and is only limited by the reduction of the $\bar\nu_e$
flux and the effect on r-processes one is willing to accept.

\section{``Conventional'' invisible channels}
\label{sec:subleading}

Before concluding, we discuss the limits on conventional (non
self-limiting) invisible cooling channels. This simple
application of the formalism set up in the previous section is
relevant to our discussion because it provides the condition under
which subleading contribution to the disappearance probability are
under control. 

Let $P_\nu(L,E)$ be the probability that a neutrino of type $\nu =
\nu_e,\nu_\mu,\nu_\tau,\bar\nu_e,\bar\nu_\mu,\bar\nu_\tau$ with energy
$E$ disappears in an invisible channel while traveling a distance $L$
and let us assume that there is no reappearance\footnote{In the case
  of oscillations, reappearance can be neglected when $P\ll 1$. The
  case of large mixing is not covered here.}. In order to obtain a
limit on $P_\nu$, we compare the consequent lepton number loss rate
with the diffusion one. The simplest case is that of a power-law
disappearance probability $P_\nu(L,E) = P_0(L/L_0)^\alpha
(E/E_0)^\beta$.  For example, the mixing with a tower of KK states
gives rise to a contribution to the escape probability which is
essentially constant in the relevant range for $L$, $E$:
$P_{\nu_e,\bar\nu_e} \sim (mR)^2$.

The neutrino number density loss rate is
\begin{equation}
  \label{eq:nloss2}
  \Gamma_\nu = \int\frac{d\vec p}{(2\pi)^3}
  \frac{\vev{P}_E}{\lambda_\nu(E)} f_\nu(E) \,,
\end{equation}
where $\vev{P}=\int^\infty_0 dx P(x\lambda_\nu(E),E)e^{-x}$ is the
average over the distance traveled of the disappearance probability.
In the case of a power law probability and within the approximations
discussed in Section~\ref{sec:assumptions}, one has
\begin{equation}
  \label{eq:gammanu}
  \Gamma_\nu = \frac{\Gamma(\alpha+1)}{12\pi^4 a_\nu}\, T^5
  F_{4+\beta-2\alpha} \left(\frac{\mun}{T}\right)\,,
\end{equation}
where $a_\nu = \lambda_\nu E^2/(6\pi^2)$. By definition of $\td$, the
diffusion rate is $\Gamma_{\text{diff}} = \nB \partial\Y/\partial t =
\nB \Y/\td$. The limit on the probability following from $\Gamma_\nu <
\Gamma_{\text{diff}}$ reflects the abundance of the neutrino type
considered, \globallabel{eq:limits}
\begin{alignat}{2}
  P_{\nu}(\lambda(\mun),\mun) &< 2\cdot 10^{-11} a_{\alpha\beta}
  \fracwithdelims{(}{)}{200\MeV}{\mun}^5
  \frac{\rho}{\rho_0}\frac{\Y}{0.3}\frac{10\sec}{\td} &\quad&\text{for
    $\nu=\nu_e$} \mytag \\
  P_{\nu} (\lambda(T),T) &< 1.5\cdot 10^{-8} b_{\alpha\beta}
  \fracwithdelims{(}{)}{30\MeV}{T}^5
  \frac{\rho}{\rho_0}\frac{\Y}{0.3}\frac{10\sec}{\td}
  &&\text{\hspace{-1.25cm}for
    $\nu=\nu_{\mu,\tau}$ or $\bar\nu_{\mu,\tau}$} \mytag \\
  P_{\nu}(\lambda(T),T) &< 3\cdot 10^{-6} b_{\alpha\beta}\, e^{(\mun/T
    - 20/3)}\fracwithdelims{(}{)}{30\MeV}{T}^5
  \frac{\rho}{\rho_0}\frac{\Y}{0.3}\frac{10\sec}{\td} &\quad&\text{for
    $\nu=\bar\nu_e$} \,, \mytag
\end{alignat}
where $a_{\alpha\beta}=(5+\beta-2\alpha)/(5\Gamma(\alpha+1))$,
$b_{\alpha\beta} = 24/(\Gamma(5+\beta-2\alpha)\Gamma(\alpha+1))$ are
numerical coefficients normalized to $a_{00} = b_{00} = 1$. In
particular, in the case of a constant electron neutrino disappearance
probability $P_{\nu_e} = (m R)^2$ we get the limit $mR\lesssim
0.5\cdot 10^{-5}(200\MeV/\mun)^{5/2}(10\sec/\td)$ that we used to
determine the conservative upper bound $\mmR\lesssim 10^{-8}\eV$.

For completeness, we also give the limit on the amplitude of a simple
oscillation probability $P_\nu(L,E) = A\sin^2(\dm{}L/(4E))$ into an
invisible channel. The limit on $A$ as a function of $\dm{}$ for the
case of $\nu_e$, $\bar\nu_e$ and $\nu_{\mu,\tau}$ or
$\bar\nu_{\mu,\tau}$ oscillations is plotted in \Fig{limits} for $\rho
= \rho_0$, $Y=0.3$, $\td = 10\sec$, $\mun = 200\MeV$ and $T=30\MeV$.


\section{Summary} 
\label{sec:summary}

The supernova plays an important role in the phenomenology of a
possible mixing between SM neutrinos and fermions propagating in the
bulk of large extra dimensions. We studied both the constraints on
such a mixing following from the necessity of avoiding unacceptable
energy loss and the implications for supernova physics. We considered
the simple case of mixing between the electron neutrino and a single
KK tower of 5D bulk neutrinos. The parameter space where interesting
effects arise is quite broad, with the size of the extra dimension in
the range $10^{-2}\eV\lesssim 1/R\lesssim 1\keV$ and the mixing mass
parameter $m$ such that $10^{-12}\eV\lesssim \mmR\lesssim 10^{-8}\eV$
and $mR\lesssim 10^{-5}$. The interplay of neutrino diffusion and
conversion into bulk neutrinos has an important role in the
phenomenology we study and requires taking into account the relevant
aspects of neutrino transport, which we did in the context of a
simplified model of the protoneutron star core.

Due to the large number of available states and especially to matter
effects, the rate at which neutrino disappearance in the bulk cools
the protoneutron star is initially dangerously high. If taken at face
value, such a rate would set quite a stringent bound on $\mmR$,
$\mmR\lesssim 10^{-12}\eV$ in the case of neutrino conversion.
However, the disappearance rate quickly reduces itself to acceptable
values. This happens for different reasons. In the region where the
MSW potential $V$ is positive, neutrinos escape in the bulk and the
lepton fraction quickly drops to the value $\Ya\simeq 0.3$ at which $V
= 0$, thus stopping the potentially most dangerous MSW enhanced
conversion before a significant amount of energy is lost.  Then, on a
much slower time scale, neutrino diffusion spoils the $V=0$ condition
and unlocks the energy frozen in what was initially the $V>0$ region.
The potential becomes slightly negative and antineutrinos start
escaping in the bulk to restore the $V=0$ condition. This feedback
keeps $\Y$ close to $\Ya$ until the temperature becomes too low to
sustain the necessary antineutrino escape rate. In the region where
the potential was initially negative, neutrino conversion never takes
place. The initial loss rate is smaller because of the much lower
antineutrino density. Still, it would be too high for large values of
$\mmR$. However, in the inner part of the $V<0$ region the
antineutrino loss stops itself by drawing $V$ up to zero before all
the local energy is lost, analogously to the $V>0$ case. In the outer
part of the $V<0$ region, there is not enough energy to reach $V=0$.
However, the conversion again controls itself. This happens this time
because the energy loss, proportional to $T^{7/2}$, becomes less and
less effective as the region cools.

The described mechanism allows to gain four orders of magnitude in the
allowed range for $\mmR$ ($10^{-12}\eV\rightarrow 10^{-8}\eV$). In the
subsequent four orders of magnitudes, the MSW enhanced conversion
would still be under control but what were previously neglected as
subleading contributions to the oscillation probability become too
large. Such contributions, as well as the limits on other
``conventional'' invisible cooling channels, were discussed in
Section~\ref{sec:subleading}. In the gained portion of the parameter
space, the implications for supernova physics are particularly
interesting. Although compatible with the \SN\ signal, the evolution
of the protoneutron star is significantly affected, especially for
large values of $\mmR$. Deleptonization and cooling take place on the
same time scale as in absence of new physics.  However, the reluctance
of the lepton fraction to get away from $\Ya\sim 0.3$ slows the
deleptonization, while the energy loss accelerates the cooling. Up to
a fourth of the energy lost by the protoneutron star goes to the bulk
but the total luminosity in all neutrino species turns out to be
approximately the same as in the standard case, at least in the first
20 seconds. Antineutrinos escape in the bulk in the effort of keeping
$\Y$ close to $\Ya$, which boosts the lepton number radiated from the
neutrinosphere. A quantitative study of the consequences of such an
enhancement would require solving the evolution of the mantle.
However, we expect it to result in an enhancement of the $\nu_e$
component of the neutrino flux and therefore of the $\nu_e$
luminosity, an effect observable in the existing neutrino detectors.
Since we expect the temperature of the neutrinosphere not to be
significantly affected, the energy deposition in a possibly stalling
shock would also be enhanced at the expenses of a smaller $\bar\nu_e$
luminosity.

The exotic phenomenology discussed in this paper represents an example
of how the standard ideas about protoneutron star evolution could be
affected in an unexpected way by the presence of new physics.

\section*{Acknowledgments}

A.R. wishes to thank M. Kachelriess, M. Keil, and G. Raffelt for
helpful discussions. This work has been partially supported by MIUR
and by the EU under TMR contract HPRN--CT--2000--00148.

\end{document}